# Impact of requirements volatility on software architecture: How do software teams keep up with ever-changing requirements?

Sandun Dasanayake | Sanja Aaramaa | Jouni Markkula | Markku Oivo

M3S, Faculty of Information Technology and Electrical Engineering, University of Oulu, Finland.

**Correspondence**
Sandun Dasanayake, M3S, Faculty of Information Technology and Electrical Engineering, University of Oulu, Finland.
Email: sandun.dasanayake@oulu.fi

**Funding information**
ITEA2 and Tekes - The Finnish Funding Agency for Innovation, via the MERgE project

**Abstract**

Requirements volatility is a major issue in software development, causing problems such as higher defect density, project delays and cost overruns. Software architecture that guides the overall vision of software product, is one of the areas that is greatly affected by requirements volatility. Since critical architecture decisions are made based on the requirements at hand, changes in requirements can result significant changes in architecture. With the wide adoption of agile software development, software architectures are designed to accommodate possible future changes. However, the changes has to be carefully managed as unnecessary and excessive changes can bring negative consequences. An exploratory case study was conducted to study the impact of requirements volatility on software architecture. Fifteen semi-structured, thematic interviews were conducted in a European software company. The research revealed poor communication, information distortion, and external dependencies as the main factors that cause requirement volatility and inadequate architecture documentation, inability to trace design rationale, and increased complexity as the main implications of requirements volatility on software architecture. Insights from software teams' awareness of the requirement volatility, factors contribute to it, and possible ways to mitigate its implications will be utilized to improve the management of requirement volatility during software architecting process.

## 1 | INTRODUCTION

Requirements are added, deleted, and modified, throughout the software development life cycle. Despite being seen as a natural characteristic of software projects, improperly management of requirement changes can have adverse consequences on project cost, schedule, and quality of the resulting product [19]. Therefore uncertain requirements are identified as one of the main factors that contribute to project failures [18]. While it's widely accepted that requirements changes are inevitable, significance of the potential consequences demands software teams to be aware of its existence, factors and impacts [42]. Multiple terms including requirements change, requirements uncertainty, requirements instability and requirements volatility are commonly associated with or related to the same phenomenon [66, 19]. Requirements volatility is selected as the preferred term to be used in this study, as it not only indicates the change of requirements but also highlights their fragile nature and the potential to be changed.

Software architecture that guides the overall vision of the software product, is one of the areas that is greatly affected by requirements volatility [8, 21]. Software architecture plays a prominent role in software development and acts as the foundation of the software systems that shapes the final outcome. Making sub-optimal architecture decisions can decrease system quality and cause problems later on [15]. Since critical architecture decisions are primarily made based on requirements at hand [31], changes in requirements may necessitate the redesign of the software to accommodate those changes, leading to an unstable software system [33]. Moreover, architecture decision making has become a collaborative team task rather than a duty of a single architect [16], therefore it's important that all the team members be aware of the possible requirement changes. Correcting shortcomings in a system at the end of its development is very complicated and far costlier than identifying and addressing them during initial design. Existing literature discusses requirements volatility in many viewpoints including project management [63, 49], coding [32], testing [40] and maintenance [57]. However, empirical evidence about requirements volatility from the software architecting viewpoint and interplay between requirement engineering and software architecture design in industrial environments is not extensively addressed.

The rest of this paper is organized as follows. Section 2 outlines the background using the related research and our motivation to conduct this study, while Section 3 describes the case study design and execution. Section 4 provides an overview of the current status of the software development process in the case company. Section 5 answers the first research question (RQ1) by describing the identified factors that contribute to requirements volatility and Section 6 describes its implications on software architecture to answer the second research question (RQ2). Section 7 takes the above discussion further by analyzing the relationships between the each identified factors and resulted implications. Section 8 discusses the factors and implications visible in each team and their readiness to handle the implications, thereby answering the third research question (RQ3). Section 9 addresses the final research question (RQ4) by discussing the means to mitigate the negative consequences and prevent them from appearing again. The possible threats to the validity of the study are discussed in Section 10 and Section 11 describes the lessons learned while conducting the study. Finally, Section 12 summarizes the findings of the study and provides the overall conclusions.

## 2 | BACKGROUND AND MOTIVATION

### 2.1 | Background

The nature of change is an integral characteristic of requirements, therefore the requirements management is referred as a process of managing changes in the requirements [64, 36]. Nurmuliani [47] defines requirements volatility as the tendency of requirements to change over time. While requirements changes can be occurred as a results of natural evolution of the user needs over the time, the actions of stakeholders in the various stages of requirement engineering process including elicitation, analysis, validation, and management can also contribute to requirement volatility [19]. Changes in market conditions, technology and regulations are some of the situations that can lead to volatile requirements and most of the time software teams have no control over them [59]. While Christel and Kang [11] group requirements elicitation problems into three categories: scoping (information mismatch), understanding (inter- and intra-group) and volatility (requirement changes), they highlighted that issues belongs to the scoping and understanding can also cause to volatility. Other important factors affecting volatility are conflicting stakeholder views and complexity of organization [24], which lead to changes in the content of forthcoming software releases. Changes to product strategy, changes to environment, scope reduction, design improvement, missing or unclear requirements, realization that original requirements were not testable and possible enhancement were given as reasons for requirement volatility from software developer' point of view [17]. The existing research studies sought factors causing requirements volatility [11], its effects on projects [66] and means to mitigate those effects [2]. Applying an iterative requirements engineering process model has been proposed as a means to address requirements volatility [39]. With the wide adoption of agile software development, the perspective on requirement changes has undergone a radical transformation as agile principles embrace changing requirement even late stages of software development and encourage to harness change for the customer's competitive advantage [9]. However, at the same time, agile principles also advices to be aware of the possible consequences of requirement changes [22].

Software architecture is the foundation that the software system is built upon. The purpose of designing the software architecture is to provide a unified vision about the system and improve understanding of its behavior. The architecture, including diagrams, use cases and semantics, reduces ambiguities and shortens the time it takes for stakeholders to understand the constraints, behavior, timing and layout for instance. Stakeholders involved in software architecture design must make various decisions throughout the software system life cycle regarding development, evolution and integration [31]. software architecture design is inherently complex, and complexity is further increased because the architecture must address various stakeholder concerns [30] that may conflict with achieving software system development goals. According to recent empirical studies on software architecture design decision-making in industrial environments, software architects use experience, intuition and other informal approaches rather than using formal tools and techniques [16, 62]. Software architecture design is considered primarily as the software architect's responsibility [14]; nevertheless, the active involvement of other stakeholders, such as software developers, product managers and customers, is crucial to better understand the criteria the architecture must meet. An important element of the architecture design process is recording architecture design rationale, as understanding the reasons behind a certain architecture design decision can be critical during software system maintenance and evolution [60]. Architectural technical debt refers to the consequences faced late in the software development process due to sub-optimal architecture decisions and trade-offs [38, 15]. Software teams accumulate architectural technical debt due to their own actions and due to external events related to natural software aging and evolution. Even though technical debt related to coding issues can be detected using various tools, architecture technical debt mostly remains invisible and grows over time [37]. Factors that contribute to accumulating architectural debt include uncertainty about requirements at the beginning of development, the introduction of new requirements during the software development process, time pressure, feature-oriented prioritization and specification issues with critical architectural requirements [41].

The requirements and architecture are considered as "twin peaks" of software development as they are equally important and interdependent. The twin peaks model emphasize on iteratively carrying out requirement engineering process and building software architecture, in parallel, in order to achieve several benefits including managing rapid changes, uncover new requirements and accommodating existing Commercial Off-The-Shelf software (COTS) [48]. Even though the traditional waterfall model leads to freezing requirements before moving into design and making hard to change architecture decisions, in reality the changes can occur in both areas and they can affect one another [12]. Quality attributes which is also referred as non-functional requirements constitute the majority of stakeholder concerns and greatly influence shaping up the architecture. The interactions between quality attributes are among the main factors that should be taken into consideration during architecture design, as architecture either allows or precludes almost all quality attributes of a system [13]. Software architecture design decisions are primarily based on architecturally significant requirements, which are critical to shaping system architecture [7]. While quality attributes take the large share of architecturally significant requirements due to their ability to affect the whole software system, it can also include functional requirements. Correctly identifying and classifying architecturally significant functional requirements is also critical for an architect to make informed decisions [45, 6].The modern iterative software development approaches facilitate close interaction between requirements and architecture and help making rapid adjustments. Terms such as continuous design [55] or evolutionary architecture[20] are used for building software architectures that can be evolved with changing conditions and it is integral part of agile software development [44]. Regardless of the adopted software development approach, the dependencies between requirements and architecture influence and constrain software development and maintenance, and shape the way software systems are built, changed and evolved [43].

## 2.2 | Motivation

Understanding and anticipating the inevitable changes occur during software development is necessary to ensure that the software product reaches the desired satisfactory levels amid the changes [50]. Despite requirements volatility is treated as a single phenomenon, in reality, different factors can lead to different types of volatile requirements and bring different implications [47]. Hence, it is important to identify the factors that cause requirements volatility, the possible impacts and the relationships between them. Even though requirement changes were frowned upon during the pre-agile era, new realities of software development demands using them as an opportunity to improve the software product [9]. However, all the requirement changes do not bring equal value to software product. While inevitable situations such as addressing evolving user needs or adopting to dynamic market conditions justify the cost of changes, it should be attempted to minimize unnecessary changes that are resulted due to process, practice or organizational issues.

Despite the major shift from building complete architecture up-front to agile friendly evolutionary architecting [20], maintaining right balance between initial architecture and its evolution during the software development process is critical to the success of software project [1]. Making excessive changes into architecture over the time can make software architecture complex, consequently costly to build and maintain [56]. It also undermines the architecture's ability to guide software development according to the overall vision. As software teams play a critical role in making architecture decisions [16], their awareness on different aspects of requirements volatility allow them to make informed decisions on accommodating architectural changes. The decisions to incorporating changes have to made carefully as they can bring both negative and positive consequences [17].

Motivated by the reasons mentioned above, this study aims to achieve two main objectives. The first objective is to study phenomenon of requirement volatility with reference to software architecture. The second objective is to understand the management of the requirement volatility during the process of architecting in software teams. The factors that cause requirements volatility, its implications on software architecture, and the relationships between them, have been identified in order to achieve the first objective. Even though the research literature provides number of possible sources of requirement volatility [47, 19, 24] as well as the possible consequences [63, 49, 32, 40, 57], this study specifically looking into the factors and implications from the point of view of software architecture. The actions taken to achieve the second objective are two fold; studying the current state of requirement volatility management and identify the possible steps to improve the current situation. Overall, this study aims to collect empirical evidence to derive fresh insights on the interaction between software architecture and requirement volatility in a modern software development setting.

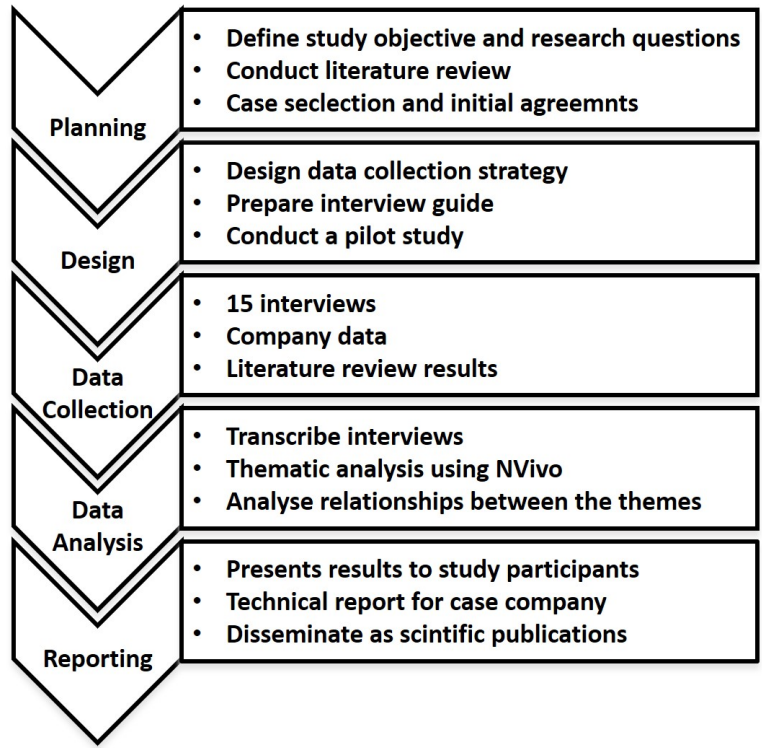

**FIGURE 1** The major steps of the case study process and the key activities carried out during each step.

## 3 | CASE STUDY

Both of the two main focus areas of this study; requirements volatility and software architecture, are heavily influenced by their operational context [61, 25]. Therefore, capturing the contextual information is essential to fully understand their behavior. Case study method is well suited for researching real life phenomena in its natural context, especially when the boundaries between phenomenon and context are not clear [65]. Hence, it was decided to select case study as the preferred empirical research approach to study requirements volatility and software architecture in an industrial setting. Since the impact of requirements volatility on software architecture in not a widely studied research topic, exploring the phenomenon and seeking new insights are the main purpose of this study. Therefore, an exploratory research strategy will be followed in this case study. The guidelines provided by Runeson et al. [52] to conduct case study research in software engineering, was used as the primary source of reference for planning and execution of this study. Figure 1 illustrates the main phases of case study process and the highlights of each of these phases. The following sub sections describe the each phase in details.

### 3.1 | Case study design and preparation

Having a comprehensive case study design that covers different aspects of the study is vital to successfully conducting a case study [52]. It was observed that the case study design not only helps to advance according to a plan, but also useful to clarify and formulate research objectives and other relevant ideas. The case study design used in this study consists of the following elements: theoretical framework, study objectives, research questions, case selection, unit of analysis, case description, methods of data collection, methods of data analysis, and validity aspects. While this list is ordered based on the general time line of the actions related to each element, it was not always taken place as a linear process. For example, formulating research questions progressed in iterative manner over a considerable period of time and the research questions were fine tuned even at very late stages of the study. Overall, case study design was constantly modified and kept up to date as the study progress.

**TABLE 1** Main themes of the interview guide and selected example questions related to each theme.

| Interview theme | Example questions |
| --- | --- |
| General information | Could you please provide a brief introduction about the company? |
|  | How are the teams organized within the company? |
|  | What is the composition of your team? |
| Software Development Process | Please describe the software development process followed in your company? |
|  | Please specify the phases of software development process followed in your current project? |
| Requirement Management | What kind of information is gathered when customers are contacted? |
|  | How do you ensure that customer request is correctly understood? |
|  | How are requirements prioritized? |
|  | What kind of requirements management tools are utilized and what are their greatest challenges? |
| Software Architecture | Please describe the software architecture design phase in your project? |
|  | How do you take consider quality attributes during the software architecture design? |
|  | What are the documents used or produced during the architecture design? |
| Challenges and Solutions | What are the main challenges relating to requirements management? |
|  | What are the challenges associated to collaborative [architecture] decision--making? |
|  | How would you improve the current way of working? |

A literature review was conducted in order to study the existing research on the subject in order build a theoretical framework. The introduction and background sections of this article are mostly based on the results of the literature review. The results of the literature review were utilized to formalize the objectives of the study and the research questions that are derived from them. This study aims to answer four research questions. The first two research questions were design to fulfill the objective of understanding requirement volatility with reference to software architecture. And the last two aim find out the management of the requirement volatility in software teams during the process of architecting. The final version of the research questions are as follows:

RQ1: What are the factors that contribute to requirement volatility?

RQ2: What are the implications of requirements volatility on software architecture?

RQ3: How well do the software teams identify the presence of requirements volatility and the factors that contribute to it, and prepare to address its implications?

RQ4: How do software teams mitigate the negative implications of requirements volatility on software architecture and prevent their recurrence ?

After formulating the initial plan, a case study proposal was sent to the case company. Since case company showed interest to further discuss about the study, the researchers and the company representatives started discussing about the practical arrangements. The company's interests are also incorporated into the case study objectives and overall case study design. Since software architecting is increasingly become a collaborative group, it was decided to use software teams as the unit of analysis. Fifteen software experts were selected to be interviewed based on the following conditions: each of them should represent a different software team, together they should cover all the major units of the case company, each of them should involve in architecting during their regular tasks and they should be available for an interview during the period of data collection. In addition to the information provide here, data collection, data analysis and validity aspects of the study are discussed in the respective sections below.

## 3.2 | Case context

The case study was conducted in a company with more than 900 employees in 25 offices worldwide. However, the product development is mainly done in three countries. The case company is a provider of comprehensive software solutions to both private and business customers and software products for private customers. The software solutions offered for business customers include a number of management tools and company services for the worldwide market. In addition, the case company provides various software products for private customers to be used on an array of devices including networking equipments, personal computers and hand-held mobile devices. The software development activities in the case company are carried out mainly in three countries.

**TABLE 2** Basic work-related information of the interviewees participated in the study.

| ID | Unit | Title | Experience (Years) | Project Type |
|---|---|---|---|---|
| SA1 | Consumer | Domain Architect | 14 | Mobile |
| SA2 | Content | Domain Architect | 24 | Cloud, Mobile |
| SA3 | Content | Software Architect | 13 | Cloud, Mobile |
| SA4 | Content | Lead Architect | 12 | Cloud, Server |
| SA5 | Content | Program Lead | 13 | Cloud, PC |
| SA6 | Content | Senior Software Engineer | 22 | Cloud, PC |
| SA7 | Corporate | Senior Software Engineer | 7 | PC |
| SA8 | Corporate | Domain Architect | 18 | Server |
| SA9 | Platform | Senior Software Engineer | 15 | Server |
| SA10 | Corporate | Software Engineer | 20 | Server |
| SA11 | Consumer | Lead Software Engineer | 13 | PC |
| SA12 | Consumer | Lead Software Engineer | 20 | PC |
| SA13 | Platform | Senior Software Engineer | 19 | PC |
| SA14 | Corporate | Senior Software Engineer | 5 | PC |
| SA15 | Technology Office | Chief architect | 15 | - |

In the case company, there are three parallel business lines: Consumer, Corporate and Content. The units are divided based on their business focus and each unit have their own financial responsibilities. There also is a horizontal business line, platform, which is responsible for providing services that are commonly shared by other three business units. The business lines operate as independent entities and within the business lines, there is a flat hierarchy of teams mostly organized based on projects. Along with the business units, a special unit, technology office, consists of technical experts who take company-wide decisions related to technical matters. Individual teams are mostly self-organizing and typically consist of four to eight people. While teams are free to operate according to their own agendas, they might have to interact and align with other teams, depending on the nature of the project. Some teams have their own architect or scrum master, but this is not the case for every team.

## 3.3 | Data collection

An interview guide, which consisted of the set of open ended questions grouped into different themes, was prepared to guide the interviews. As shown in (cf. Table 1), the interview guide[1] consisted of the following themes; a general information, software development process, requirements Management, software architecture, and challenges and solutions. The background questions covered the general information such as the experience of the interviewee, typical project and team composition as well as company description. The software development process theme covered the overall development life cycle. Rest of the themes were designed to go into a deep discussion about requirement engineering and software architecture design activities. The final part targeted, the various challenges faced by the interviewees and the possible solutions from their perspectives. A pilot interview was conducted to test the interview guide as well as the interviewers approach. The participant of the pilot interview had a long history working in industry as a software developer and a software architect. In addition, the feedback received from the company representatives was also used to improve the interview guide. The reviewers of the questionnaire and the pilot interview participant did not participate in the actual case study.

Fifteen semi-structured, thematic [26] interviews were conducted with software experts as the primary method of data collection. All the interviewees acted as software architects in their respective teams even though not all of them are titled as architects (cf. Table 2). The interviewees were selected to represent all five units in the company's technical organization. Interviewees have different levels of work experience, are active in various projects. The experience given in (cf. Table 2) reflects the years of experience in software development in industry. The interviews lasted from one to two hours and were conducted by two researchers. Twelve of the interviews were conducted face to face (F2F) at case company sites and three via Skype. Interviewees were provided in advance with the list of interview questions, which was used as a guide, and more detailed questions emerged during the interviews.

---

[1]`https://dasanayake.com/jsep_interview_guide.pdf`

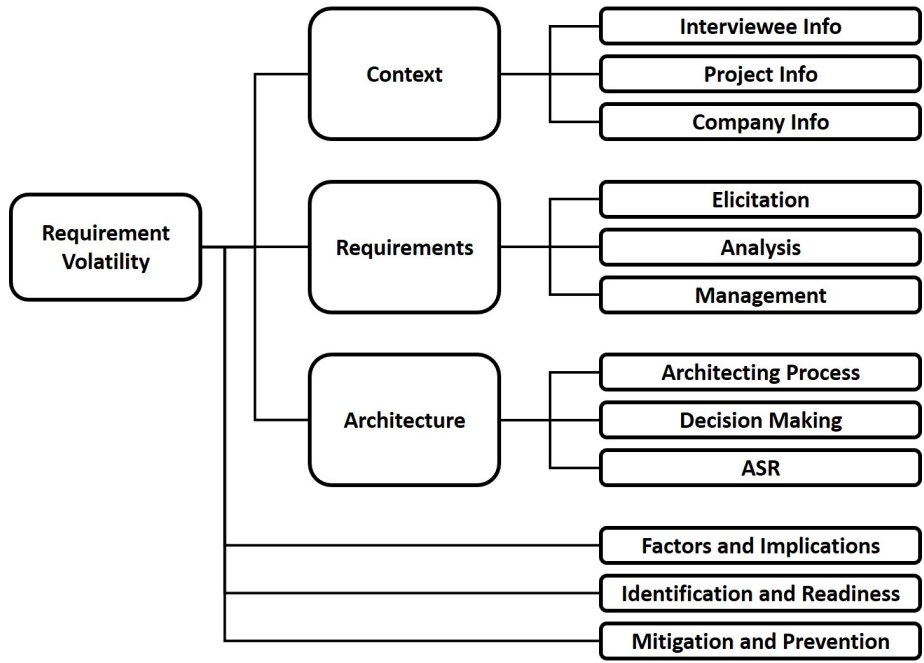

**FIGURE 2** The major themes and sub themes used during the data analysis.

## 3.4 | Data analysis

The interviews were professionally transcribed and fed into NVivo, a qualitative data analysis tool, as the starting point of the data analysis. None of the respondents was a native English speaker. Therefore, it has been necessary to correct some minor language errors to ensure a proper message. The researchers went through each of the transcribed interviews and labeled the relevant information using a predefined set of themes based on the interview guide. Initially, there were the limited set of themes based on the research questions and as the process going forward, new themes were emerged and the sub themes were also created. The themes used during the data analysis are shown in Figure 2. The themes are not mutually exclusive and the same data can be labeled with multiple labels belong to different themes.

A strict privacy policy was followed that described all necessary elements of the case study while protecting the integrity of the company and individuals [5]. Interviewees' identities are protected by, for example, using aggregated information instead of presenting interview excerpts and by avoiding use of corresponding IDs in tables. The topics represent the aggregated viewpoints of the respondents throughout the interviews. The issues mentioned only by a single interviewee have not been included in the results, as they were considered not expressing shared understanding among software architects. Quotes from interviews are attached with the relevant results as quotes provide insights for collected empirical evidence. However, to protect the integrity of the respondents the quotes are not labeled by the respondent ID.

## 4 | SOFTWARE DEVELOPMENT PROCESS IN THE CASE COMPANY

Software development practices in the case company are quite informal and vary from team to team. The company has a long history with various lean and agile software development approaches. At the moment, there is no company-wide software development approach. Software teams are free to select their own approaches unless there are specific restrictions such as customer preferences. Software teams tend to create their own approach by selecting and combining various agile and lean practices, including following sprints, maintaining product backlog and using Kanban boards. Although interviewees' have responsibility for software architecture design, their involvement in requirements engineering is limited to occasionally providing expert opinion. Thus, interviewees' understanding about requirement engineering is not as extensive as is their knowledge of software architecture design.

In the case company, the life cycle of each product is managed separately. The team that has the responsibility of managing product is referred as "Product Management". They make the decisions about the product from its conception to delivery. They are also responsible for managing the maintenance and the evolution of the product. That includes managing different variations of the products across different platforms such as desktop, mobile, and web, as well as services supporting that products. In order to do that, product management create the product road maps and figure out the high-level product requirements. However, the direct interactions between the product management and the actual software teams that develop the product is quite limited. The product owner is the one who act as the link between the software team and the product management. In addition to conveying product management's vision and requirements to software team, product owner works together with the team to make sure that the developed product fulfills that requirements. From the point of view of the software team, product owner is the internal customer who represent the needs of the actual customer.

### 4.1 | Requirements Management

Software architects are not directly involved in customer requirement elicitation and analysis. In the case company, elicitation is accomplished by using techniques such as focus groups, beta testers, and direct customer communication. Occasionally, customers' ideas are expressed at so abstract a level that they can hardly be translated as requirements. On the other hand, it is possible that requirements will be stated as full-scale technical specifications. Customer needs are clarified during the requirement analysis. The product owner is the link between the software team and product management. In the case company, the term "Product Management" is used to refer the team of that make the decisions about the product road maps and figure out the high-level product requirements. The product owner has the responsibility to communicate with the product management to identify the product requirements and then convert them into user stories together with the software team and monitor their progress. For the majority of interviewees, the product owner is the sole connection point to requirements elicitation and analysis phases. During the clarification process, software architects are occasionally consulted to define the technical feasibility of requirements and to choose the best implementation solution. At this phase, requirements volatility factors include changing customer needs and evolving technological understanding.

Jira - the issue and project management tool is the main medium for documenting requirements, which usually are expressed as features or backlog items. In addition to the primary Jira, legacy and team-specific tools and sticky notes are used to communicate tasks, store customer information and document decision rationales. The level of details in information on requirements varies depending on a product and a team. Sometimes, only a feature name is recorded, but at the other extreme, descriptions include even the contact information of the relevant technical specialist on the customer side. In the case of private customer products, requirements are created by experts based on a foreseen market. Usually, the creator of a requirement is recorded, but sometimes it is not known whether the requirement originated with a customer or an internally identified technology gap. Interviewees stated that they sometimes needed more technical details or contextual descriptions to be able to choose the best implementation alternative. Big corporate customers may have strict requirements about formal documentation to be delivered to the customer. Interviewees pointed out that a requirement description is always a compromise between level of detail and time available for the task.

The product management is responsible for prioritizing requirements by taking into account the factors such as company strategies and the importance of the customer. The dominant factor considered while prioritizing requirements is the customer. The bigger the customer, the higher the priority of its requirements. Even though some features are technically feasible and could contribute improving the product quality in the long run, it may be very difficult to reject a request of a customer, especially if the customer is very important to the company. Other factors taken into account when setting requirement priorities include development cost, feature size, product road map, criticality and external audit results, which are publicly available and used to rank the solution providers in the domain. Interviewees were involved in requirement prioritization by proposing product improvements and project scoping meetings. Most interviewees mentioned having faced challenges with changing requirement priorities. As the backlog is updated frequently, changing priorities contribute to requirements volatility.

### 4.2 | Software Architecting

Since the majority of the teams follow agile and lean approaches, the design and implementation are done iteratively, leading to a shorter design phase than in the traditional waterfall approach. The software architecture design process typically starts with backlog review meetings between the team and other relevant stakeholders. The objective of these meetings is to reach a consensus what needs to be developed to fulfill requirements. While the team is generally represented by senior members during the initial meeting, it is possible that the whole team is involved from the beginning. Once the basic architecturally significant requirements are understood, the team creates a design proposal, which is delivered for review. The review is done at different levels, depending on the scale of the project or its dependencies to other products. Once decisions are made, the team is free to begin development and has the flexibility to make minor changes to the design. If the design must be altered considerably, the evaluation of the alterations is escalated. software architecture designs and decisions taken during discussions at various levels are recorded using several methods. Even though the interviewees claimed that they have maintained some type of design documentation, attention to documenting design appears to be inadequate. The majority of teams use tools such as Wiki or Jira instead of traditional design documents to store their architecture decisions.

## 5 | ORIGINS OF REQUIREMENT VOLATILITY - CONTRIBUTING FACTORS

The first research question (RQ1) is answered in this section by describing the factors that contribute to requirement volatility. Ambiguous requirements, changing user needs, dynamic business environment, external dependencies, information distortion, ineffective communication and change of personal are the main factors that are identifies as contributors to requirement volatility. Based on the nature of each of these factors, it's possible to recognized three different groups of them. The first group is the factors related to information management. Ambiguous requirements, information distortion and ineffective communication can be grouped into that as they all are related to inability to obtain necessary information. Then another two factors, changing user needs and dynamic business environment can be places into a single group as the factors related to operational business domain. Finally, external dependencies and change of personnel can be place in the uncontrollable factors group as the software team doesn't have any control over them. While some of these factors are directly responsible for bringing changes to existing requirements or adding new requirements, the others' role in that regards is not visible at the first glance. Rather than changing requirements at a given point, they can degrade the quality of requirements over a long period time and gradually increased the requirement volatility. The descriptions of each identified factor and evidence of their presence are given below.

### 5.1 | Ambiguous requirements

Most common reason for ambiguity of requirements is inadequate or missing descriptions of backlog items. Often, features or backlog items lack detailed information; for example, a backlog item may have only a name, but no one knows why the item is there. The tool includes a customer acceptance criteria field as part of the requirement description. Most of the time, something is recorded in this field. However, the description is often a couple of lines of text at an abstract level. This means that architects and testers must guess what must be fulfilled.

*"It [description] can be just couple lines of text and that's all and we need to guess what shall we do…. quality engineers always complain about it because they don't know how to test because it's not so clear how it should work."*

### 5.2 | Changing user needs.

The case company provides multiple software products for various customer groups and frequently comes across changing customer needs. As most of the requirements for private customer products are decided within the company, corporate customers are the main source of requirements changes. The long-term business relationship between corporate customers and the company makes it difficult to refuse to adapt to changing customer needs.

*"Well, since we are doing this project with, constantly changing requirements. I don't see much chance for improving the process because we are just, basically adapting. And not planning ahead."*

### 5.3 | Dynamic business environment.

The company operates in a dynamic business domain and must adjust its strategies for accommodating development in that domain to stay ahead of the competition. The severe market situation requires constant changes in requirement priorities. As most of the company's private customer products run on smartphones where the operating systems are highly fragmented and subjected to frequent changes, the company has to make frequent changes to their products to accommodate those changes.

*"This list we see for quarter is something that we can work on. Whatever in future is, at least, that is subject to change because market change, situation change and stuff like that. So we wouldn't know."*

## 5.4 | External dependencies.

The company is structured along business lines, each of which runs its projects independently. However, sometimes delivering a solution requires collaboration among teams from different business units. For example, a team that has developed a mobile application might have to interact with teams that have developed the same application for different platforms and with teams that provide server-side support for those applications. In this situation, changes in requirements in one unit lead to changes in another one.

*"When we have external dependencies on the teams in, especially if they are another location, it's sometimes quite hard to make sure that everything happens in time."*

## 5.5 | Information distortion

Majority of the software teams in the case company have no interaction with the actual customers. They receive the requirements from the product owners or some other source within the organization. In some cases, information originated from the customer has to flow through several different external and internal units before reaching to the development teams. The information are subjected to be distorted in each of the handover and the greater the distance between the customer and the development team the higher the level of possible distortion.

*"So I don't contact the customers directly. Sometimes they ask something via emails, but they are never end-users. They are maybe our operator customers, or our salesmen that ask some questions but, it's not in my interface that I'm directly."*

## 5.6 | Ineffective communication

As the case company has a globally distributed customer base, multiple development sites and virtual software teams, it is challenging to communicate requirements. Communication issues may begin during the elicitation and customer negotiation phases. Typically, this is due to the fact that the terminology and semantics differ between customers and developers. These differing domains bring to the table different terminology and concepts. Later in the process, software teams may face language barriers and cultural differences that pose communication challenges. Communication issues are present, even though the company has tried several supporting tools and approaches to improve communication both within the company and with customers.

*"And we often need clarification all the way from customer, but we do have this kind of feedback cycle from us to the customers, so that we can find out what exactly is wanted. Because that's really not that clear always."*

## 5.7 | Change of personnel

Changes within the software team and other stakeholders that are actively involved in software development process contributes to requirement volatility. As mentioned above, many software teams state that the requirements descriptions are ambiguous and not well documented. Hence the software architects have to rely on the individuals with the domain knowledge to acquire and maintain this information. However, constant change of personnel leads to loss of this knowledge.

*"I've been three years in this process and this is the third product owner. So, they don't know the product. They don't know the platform and they don't know what we can implement."*

# 6 | REQUIREMENT VOLATILITY IN ACTION - IMPLICATIONS ON ARCHITECTURE

This section provide answers to the second research question (RQ2). Interviews revealed several implications of requirement volatility that adversely affect software architecture. They are time and resource management, stakeholder synchronization, architectural technical debt, architecture documentation, tracing design rationale, maintaining quality attributes and architecture complexity. While some of these implications directly affect the architecture itself and the others affect the process of architecting, which deals with conceptualizing, building, maintaining and evolving the architecture. Even though we only considered the implications on the architecture in this study, the issues related to time and resource management, stakeholder synchronization and maintaining quality attribute cam negatively affect whole software development process and the resulted artifacts. It should be noted that requirements volatility is not the sole reason for the identified implications on software architecture as there are many other reasons including contextual, process and organizational factors affect their severity. However, requirement volatility was highlighted as one of the main contributors. The subsections below describe each of the identified implications in details.

## 6.1 | Time and resource management

Typically, requirement clarification takes considerable time in the case company and because of that, architects receive requirements very late in the project. This causes challenges in scheduling both on the team level and the organisational level. On the team level, the later the architects receive the requirements, the less time they have to design architecture. Most of the software teams use prototyping as one of the main technique to decide between the possible architectural alternatives. Since change in requirements make previous architecture decisions redundant, the team has to put an effort to make a new round of prototypes. In the organizational level, the management creates development plans quarterly. The aim is to have two forthcoming quarters planned to provide an overall idea about what should be achieved in six months. However, quarterly plans are subject to change. Often, when a quarter begins, the schedule applies for only a couple of weeks, and then the content must be re-planned. In the worst cases, priorities change daily, in which case, architects have no choice but to work on the item at hand and then take the next one on the backlog list, which might be different the next morning.

*"I think the main problem is that, we get these, requirements are coming so late, that we need this feature. We have so short period for, making designs or, basically some is skipped. It's just implementation and change the design during implementation."*

## 6.2 | Stakeholder synchronization

Stakeholder dependencies are one factor causing requirements volatility, which, in turn, causes synchronization issues. Interviewees noted that sometimes they are unable to deliver products on time due to delays in other units. Beside inter-unit synchronization issues, teams in the same business unit but at different development sites have synchronization challenges caused by lack of physical proximity. Synchronization needs and development dependencies also influence the tools used. Jira was used for project management, requirement descriptions and bug fixing. These activities require very different tool functionalities, which are supported only partially. For example, maintaining the backlog within a project and following feature development work well, but cross-project management, such as moving a feature from one project to another, is not supported.

*"Probably the most complex thing is that, we need to somehow synchronize the requirements between the different teams and, that's why, having some leadership team would be beneficial because they would synch up together what they are gonna do, what resources they have, how they would transfer their stories between the teams and et cetera."*

## 6.3 | Architectural technical debt

The company's requirement prioritization criteria are strongly business driven, favouring market needs over architectural considerations. Overlooking architectural aspects when prioritising requirements accumulates architectural technical debt. As architects are overwhelmed with volatile requirements, they are not necessarily able to find out the optimal architectural design choices. Specifically, prioritising functional requirements over non-functional requirements is a major issue, as non-functional requirements are the majority of architecturally significant requirements, and neglecting them leads to sub-optimal architecture design.

*"I think the biggest driver usually for getting something prioritized really fast is money. So if the customer is big enough and the expected income is big enough, [case company] will run through hoops..., for smaller customers even if the smaller customer is asking for something that makes much more sense. So I think that, the primary driver is economic. So instead of doing feature development, we kind of get overridden by the [business customer] deliveries all the time, because that's where the money comes from."*

## 6.4 | Architecture documentation

The interviewees understand the importance of documentation as it helps them to preserve and share knowledge, specially during the later stages of software development life cycle. While many interviewees claim that they have started some form of documentation at the beginning of the project, later they find it difficult to maintain design documents as they require frequent updates due to the volatile requirements. In most cases, documentation is limited to information stored in Jira and code comments.

*"If we need some sort of documentary, because there is a thing that, at this state of a project, document itself does not make any sense. Because technical requirement or the feature requirement are changing so often, then, as the document even been written, it is just, becomes obsolete. That's why we are not putting them into the Wiki and saying that this is our technical way how it's supposed to be. Instead we are trying to communicate, and agree on some long or short term plan."*

## 6.5 | Tracing design rationale

As every element of software architecture is a result of a decision made to address an explicit or implicit requirements, it should be possible to trace back to related requirements during the later stage of the software development process. Recording design rationale, the reasons behind a given decision while making an architecture decision is a vital activity that facilitate the traceability. However, requirement volatility not only discourage software teams from recording it but also muddle the information that are already recorded. Even though Jira supports current need to some extent, there are major issues with outdated and unstructured information, broken links and inability to trace decision rationale.

*"So usually the reasoning [behind design decisions], happens, it's kind of like corridor discussion where we have a meeting where we talk about it and then we report what we decided to do in the Jira items but of course there's lot of stuff that we miss, so we don't really document the why, we document the what. And of course these discussions in the meeting, kind of contain the why also but if you are not in the meeting then that information is not available, usually."*

## 6.6 | Maintaining quality attributes

Even though there are large variations in emphasize on quality attributes between the different teams, every interviewee stated that they always consider quality attribute to some extent. Software teams are facing many issues related to integrating quality attributes as they are stated implicitly and ambiguously among the requirements. Most of the time it's up to the software team to define, integrate and maintain quality attributes. Maintaining quality attributes has increasingly become difficult when the volatility of requirements are greater. Since the quality attributes cross-cut through many parts of the software system and they are intertwined with many other requirements, every change in requirements can potentially impacts quality attribute.

*"Maintaining that kind of system is a lot of work and quite often someone forgets to check the results after a change. Usually it takes a long time for us to discover that there has been a performance regression, unfortunate as it is."*

## 6.7 | Architecture complexity

When future requirements are unclear, software architects have two possible paths to design the architecture. One option is building a full scale initial architecture that is flexible enough to accommodate most of the future requirements. In this case, the complexity of the architecture is high and it also requires more time and other resources to implement it. While it helps to easily address the future requirements, building and maintaining a complex software system is an error-prone process. The other stated option is keeping architecture simple and evolve it as time progress and new requirements received. While most of the teams tend to follow the second option as it's well aligned with agile principles, there are some teams that opts to follow the first option.

*"Our solution is quite flexible. In the sense, that means they are quite complex. Because you should be able to enable different features whenever our operator customers wants"*

# 7 | RELATIONSHIPS BETWEEN FACTORS AND IMPLICATIONS

Even though various different types of changes can be added under the umbrella of requirement volatility, all the changes do not have similar implications on software project. For example, making modifications to different elements of requirements brings different results. While the impact of changing the priority of a given requirement might be limited (ex: *"Our priorities change all the time, sometimes daily. So, we basically get the next item and work on it."*), some other changes in the requirements can cause a larger impact (ex: *"We are kind of suffering from the complexity because, of change in requirements, not because of bad architecture."*). Hence it's important to identify different types of requirement changes cause by each factor and what kind of implications they can have on architecture. It allows software teams to trace back to the factors that cause certain issues and prevent them from happening.

Figure 3 illustrates the relationship between the identified factors of requirements volatility in the case company and the implications on software architecture. The links between the factors and the implications are constructed based on the statements made by the interviewees. The aim of the diagram is not to illustrate the frequency of each relationship, but to illustrate the types of implications occurs as a result of each factor. It doesn't also represent the severity of the impact as it is very context dependent. Even though changing user needs affects only two of the implications, repeated frequent changes in large set of user needs, can have a severe impact than a set of ambiguous requirements that are not very important to the software system.

Based on the illustrated relationships, ambiguous requirements and dynamic business requirements are the most capable of bringing different types of implications. Even though both of those factors negatively affects architectural technical debt, architecture documentation, maintaining quality attributes, and architecture complexity, the ambiguity impacts them due to missing clarity while dynamic business environment adds the time pressure. The ambiguity forces software teams to aim into a target that is not entirely clear and the dynamic business environment keep moving the target without allowing them to settle in properly. Change of personnel contributes to four different types of implications. Among those, two of them; architecture documentation and tracing design rationale, are the results of the unshared knowledge left with the old team member. The time and effort require for a new team member to settle in, can cause issues related to time and resource management, as well as stakeholder synchronization. Information distortion and ineffective communication both causes similar issues due to missing or unclear information sharing. External dependencies make synchronization difficult and lead a complex architecture that facilitates all the dependencies. Finally, changing user needs, creates time pressure even though it doesn't bring large set of consequences similar to dynamic business environment.

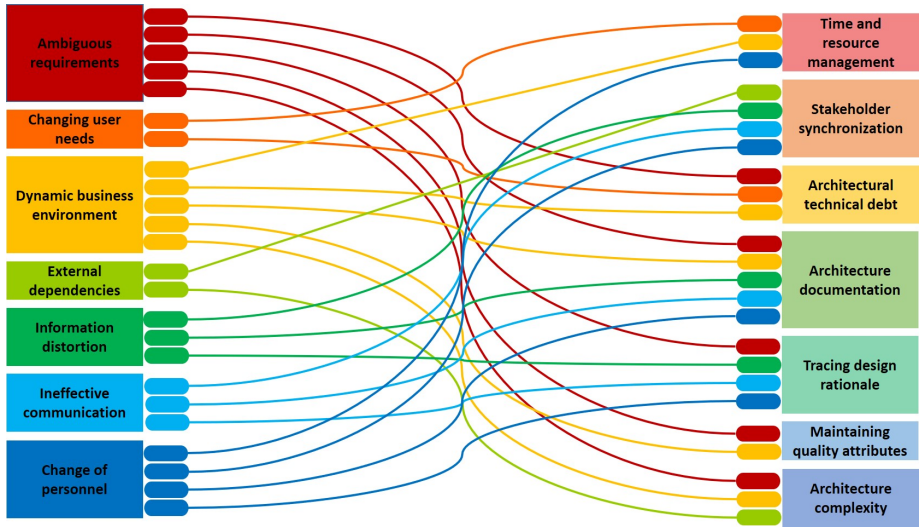

**FIGURE 3** Relationships between the factors that contribute to requirement volatility and the implications of requirements volatility on software architecture.

From the architecture point of view, architecture documentation can be degraded due to the requirement volatility caused as a result of all the factors stated above, except external dependencies. In addition to documentation, both tracing design rationale and stakeholder synchronization are affected by similar factors: Change of personnel, ineffective communication and information distortion, collectively. All of these factors are related to handling information and missing information and the negative implications are brought to the processes that are dependent on reliable information.

## 8 | DIAGNOSING REQUIREMENT VOLATILITY - IDENTIFICATION AND READINESS

The goal of this section is to answer the third research question (RQ3) by identify the presence of requirement volatility, the factors contributed to it, its implication on each software team and their readiness to handle those implications. Despite being part of the same case company, each team has different combination of characteristics such as project type, stakeholders and team culture that differentiate it from the other teams. Hence, impact of requirement volatility on each team is different. While identification of possible factors and consequences of requirement volatility in each team is important, it alone doesn't ensure that software teams are ready to face the implications and mitigate the challenges posed by them. Table 3, contains information on factors of requirement volatility and implications of requirements volatility and readiness of each team to handle those implications.

### 8.1 | Identifying the requirement volatility, factors and implications

Every software architect participated in the study has recognized requirement volatility as a part of the daily software development process of their respective teams. Based on the information discovered, a handful of factors and implications dominantly visible in the majority of software teams. Ineffective communication is the most dominant factor of all. 12 out of 15 interviewees admitted that communication is one of the main areas that should be improved. External dependencies and information distortions are closely followed it as other concerned

**TABLE 3** Presence of the factors that contribute to requirement volatility and the implications of requirements volatility on software architecture, in each team and their readiness to address them.

| ID | Sources | | | | | | | Implications | | | | | | Readiness |
|---|---|---|---|---|---|---|---|---|---|---|---|---|---|---|
| | AMB | CNG | DYN | EXT | INF | COM | PER | TIM | SYN | ATD | DOC | TDR | QUA | CPX | |
| SA1 | X | | | X | X | X | | X | X | X | X | X | | | Positive |
| SA2 | | X | | X | | | | | X | | | X | | X | Positive |
| SA3 | X | | X | X | | | X | | | X | X | X | X | | Neutral |
| SA4 | | X | X | | | X | | X | X | | X | | | | Negative |
| SA5 | | | | X | X | X | | | X | | X | | | X | Neutral |
| SA6 | X | | | X | | X | | | | | | X | | X | Neutral |
| SA7 | X | | | | | X | | X | | | X | X | X | | Neutral |
| SA8 | | X | | X | X | X | X | | | | X | X | | X | Negative |
| SA9 | | | | X | | X | | | | | X | | | X | Positive |
| SA10 | | | | X | X | X | X | | | | X | X | | | Neutral |
| SA11 | | | | X | X | X | | X | | X | X | | | | Neutral |
| SA12 | | | X | X | | | | X | | | | | X | | Neutral |
| SA13 | X | X | | | X | X | | | | | X | X | X | X | Negative |
| SA14 | | | X | | X | X | X | X | | X | | X | | X | Negative |
| SA15 | | | X | | | X | | X | X | | X | X | | | Positive |

AMB - Ambiguous requirements; CNG - Changing user needs, DYN - Dynamic business environment, EXT - External dependencies, INF - Information distortion, COM - Ineffective communication, PER - Change of personal, TIM - Time and resource management, SYN - Stakeholder synchronization ATD - Architectural technical debt, DOC - Architecture documentation, TDR - Tracing design rationale, QUA - Maintaining quality attributes, CPX - Architecture complexity areas. Collective those three areas appears to be have a considerable impact on the teams. However, since they are mostly interdependent, addressing one of the issues can positively improve the other aspects also. For example, improvement in communication can reduce the information distortion and it also helps to manage external dependencies in a better way. Even though requirement ambiguity is considered as one of the main contributor to requirement volatility [32], only a handful of the teams in the case company have mentioned that as a possible concern. This can be attribute to the strong presence of the role of product owner as the middle man between the software team and the rest of the stakeholders. ("*Thinking from our angle there is no challenges. Because, we get them delivered on a silver platter because the product owner just makes a judgment call. The actual difficulty then lies with the product owner who has to fit the actual business objectives with the exact right, feature content, balancing.*"). Even though few of the interviewees mentioned about changing user needs and dynamic business environment as possible factors, majority of the teams in the company do not affect from these situations and work on based on their own schedule. Specially, those teams who work on consumer project have liberty to work on their own while teams from corporate and content business are affected by the nature of operations of their corporate customers.

Among the implications to software architecture, documentation is the area that is greatly affected, followed by tracing design rationale. Similar to the situation in the factors, both of these are also closely related. Documentation appears to be an issue in large majority of the teams. While some of them are concerned about it, others consider it as the norm regardless of the consequences. After that, complexity is an issue for considerable number of the teams and the main reason appears to be integrating a large set of features into the software. Time and resource management and synchronization between the stakeholders are present among little above 1/3 of the studied teams. Architecture technical debt and maintaining quality attributes are considered as implications by only a small number of teams. Overall it appears that managing communication, collaboration and documentation are the most important aspects for the software teams dealing with requirement volatility.

## 8.2 | Readiness to handle the implications

The readiness of each group to handle the implications of requirement volatility, is categorized into three groups: positive, neutral and negative. The software teams who actively anticipate requirement volatility related challenges and are confident that they can handle them, are categorized as "Positive" in the readiness level. As example, the following remarks are made by the members from some of the teams belongs to this category.

"I'm okay with the changes, even if they come at late point."

"I usually try to make things a bit more abstract than they actually need to be so all these changes have so far, haven't required any re-writing of code. Everything has been rather smooth. Basically, the idea is to support flexibility, even at the cost of simplicity. Because, we don't know how the requirements are going to change. So we are just trying to adapt. Make sure that the application can adapt."

The readiness level "Neutral" is assigned to the teams that don't necessarily see requirement volatility as an issue for their team. Even though requirement volatility factors and some of the possible consequences are visible, they don't consider those implications as results of requirements volatility. This view can be justified as there are many factors other than requirements volatility that brings similar implications. It is also possible that they accept those consequences as a part of the normal software development process and don't see a need to deal with them.

"Requirements that come from product management, they're pretty high-level. I don't think that they can be concrete or have very, like exact numbers. I don't remember that they usually have even any kind of exact performance figures or anything like that. Those are usually something that, we then, try to come up with ourselves."

"I have almost everything. What I don't have I can gather by using daily contacts with people here and there around the business line."

The teams who are unable to take any actions despite suffering from the negative consequences of requirements volatility are categorized into "Negative" readiness level. This can result due to various reasons including the reasons that are beyond the control of the respective team.

"We do have a long list of prioritized items that we should work on. In practice, sometimes requirements come on the fast lane and then we just need to work on those because some corporate customer is requiring this and that and then we just need to work on that even though it's not a priority from product perspective or from engineering perspective, but since it's just so important from sales perspective that we just have to do it."

"Normally requirements are prioritized by product owners. And also, sometimes there are escalations from support for instance and some firefighting. Especially our team maybe, actually, we can't use Scrum basically because our priorities change all the time. We basically get the next item and work on it."

Closer to a half of the studied teams (7/15) have a neutral attitude towards requirement volatility. Even though they see some consequences, they don't consider it as a pressing issue. The rest is divided equally among those who think that they are well equipped to handle the requirement volatility and the other who are unable to solve the requirement volatility despite being affected by the negative consequences of requirements volatility.

## 9 | ADDRESSING IMPLICATIONS OF REQUIREMENTS VOLATILITY - MITIGATION AND PREVENTION

This section answers the final research question (RQ4) by discussing how software teams mitigate the negative implications of requirements volatility on software architecture and prevent their recurrence. The discussion focuses various organizational, process and tooling changes to reduce the severity of the implications and eliminating the factors of requirements volatility to prevent or minimize occurrence of requirements volatility, in relation to existing scientific knowledge in the light of the empirical data gathered from the case company. This two pronged approach not only helps to overcome the implications of requirement volatility related challenges in short term, but also brings long term benefits to the software teams by eliminating the primary source of these implications.

### 9.1 | Mitigating the implications on software architecture

As previous sections have described, the correlations between the factors and the implications can be identified. How ever, the evidence from this study is not sufficient to suggest any causal relationships between them. Therefore, there is no definite solution to address each identified consequence. While eliminating all the possible factors appears to be the obvious solution, the impact should be carefully evaluated as it can lead to unexpected results.

Software teams should be encouraged to improve the documentation practices and recording design rationale by providing adequate tools [60]. In addition, a company-wide methodological approach to recording design rationale and maintaining necessary documentation should improve the quality of documentation and traceability of design rationale. As scheduling issues appear at the team level and at the organizational level, addressing them must be undertaken at both levels. At the team level, the situation could be improved by, for example, assessing the suitability of the elicitation techniques used and the adequacy of the requirement information collected. According to Hickey and Davis, an elicitation technique should be chosen based on the problem, solution and project domain characteristics as well as known requirements [27]. On the organizational level, one solution could be to include a sufficient buffer [23] for planned releases as a response to requirements volatility. According to feedback from interviewees, interaction among team members located at various sites is not adequate, despite using various communication tools. When it comes to distributed teams, just maintaining work communication among team members is insufficient. The performance of distributed teams is affected by networking within the team and trust among members [58].

Lack of visibility among business units was mentioned constantly during interviews as hindering synchronization among business units. While separations among business units may be necessary to organizational management, they cause several negative results in software architecture design, the main one being the possibility of duplication of work, as the teams are not aware of each other's work. Considering the amount of human resources and talent in the case company, there are good opportunities for knowledge-sharing among engineers. Even though the technology council and company-wide steering group can prevent the large-scale duplication of effort, work still can be duplicated on the micro-level. Closer interaction among software teams in various business lines will facilitate the identification of resources suitable for a given task and, hence, get it done more efficiently. Since individual business lines evaluate their own performance, collaboration with other business lines might not be high on their agendas. However, in the long run, business lines and the company as whole can benefit from a transparent approach.

Taking the views of software teams into consideration during the prioritisation process can contribute to reducing architectural technical debt considerably. Since it is the software architects' responsibility to recognize the architecturally significant requirements that include non-functional requirements and their effects on overall software system architecture, architects are in the best position to identify factors causing debt and manage them to minimize accumulation of architectural technical debt [62]. In the context of iterative software development, addressing architectural technical debt as a separate backlog item can ensure that it is addressed properly, as otherwise, the cross-cutting nature of non-functional requirements makes it difficult to address them properly at any given point [46]. Improving traceability across the software development process is important to understand the implications of changing requirements for software architecture and assessing possible architectural debt [34].

The existing tool chain of the case company, including Jira, Wiki and a version control system, can be used as a pragmatic solution to help mitigation the impact of requirement volatility. Since the tool chain is already in use, it doesn't cause any additional burden of tool maintenance and support. Among those tools, especially, the proper use of Jira can help to mitigate many identified issues as it was widely used by the software teams in the case company throughout the software development process. Jira facilitates structured recording of requirements, refining them into different hierarchical levels such as epics and stories, and linking them based on various dependencies. Those functionalities can be used to make sure that requirements are properly collected and documented. The tool can also help maintaining traceability as the users can be navigate through various steps of software development process to link design and development tasks to corresponding requirements. The burned down charts and other resource management mechanisms available in Jira can help software teams to have some control on the issue of time and resource management. In addition, the tool chain can be also used to increase the visibility among different stakeholders as it allows them to refer to the information in the wiki, monitor the progress via Jira and the status of the existing software product using the version control system. However, to get the maximum benefit from the tools, it is important to educate engineers and provide necessary training. In addition, filling gaps in the existing tool chain or introducing a new tool chain that provides end-to-end tool solutions would help address identified challenges.

## 9.2 | Preventing the recurrence of the implications

Introducing advanced collaborative and communication mechanisms can improve the communication among distributed software teams [4] as well as customer communication issues [35]. At the same time, using multiple communication channels rather than a single communication channel can help avoid misunderstandings caused by cultural and language differences [54]. In the context of the case company, most of the time the information from the customer should pass via several stakeholders before reaching the software team and vice versa. Maintaining a direct communication between the software teams and the customer will bring several benefits including minimizing information distortion, reducing the waiting time and taking prompt actions, in addition to improving the quality of communication. A strong, mutually beneficial relationship between the customer and the development team is crucial to successfully managing changing customer needs [28]. While this helps to understand the customers' true needs and derive well-defined requirements from the beginning, it also allows developers to communicate the consequences of accommodating changes rather than blindly accepting them. Approaches to mitigate the effects of changing customer needs include eliciting gaps in requirement changes [51] and reusing existing requirements to identify the gap between elicited requirements and true user needs [3].

As the software teams in the case company greatly depend on each other's work, their requirements are interconnected. Therefore, it was alarming to observe that identifying the dependencies between the requirements of various stakeholders sometimes was more wishful thinking than practice. Managing requirement dependencies is acknowledged as a challenging task [10] that may be addressed by supporting impact analysis [63]. As the product owners role is already proved to be very useful in coordinating the software teams with the other stakeholders including the internal teams in the case company, one approach could be to establish a product management team across business units that would have overall responsibility for managing projects that span business units, including resources, scheduling, priorities and such.

Interviewees provided contradictory opinions about how much information is available to them. On one hand, it was reported that at the beginning of a project, a significant amount of time is spent negotiating technical feasibility and clarifying actual needs, the intended behavior of the product and dependencies with other features. On the other hand, it was noted that if the available information is not too detailed it leaves room for creativity and allows discovering the best technical solution. Missing requirements early in the process causes the costliest fixes later in the development process [47]. The starting point for addressing requirement uncertainty is to evaluate what information is crucial to whom, why and when. This should be stated explicitly in the information fields provided for describing the requirement. Unnecessary default requirement fields should be removed. However, this is not sufficient, since the quality of the descriptions depends on the expertise of the writer. According to interviewees, POs or marketing personnel do not have enough technical understanding about the product to write detailed requirement descriptions. In addition, it would be a waste of time to write an extensive requirement description just to find out later that the requirement is not technically feasible. The impact caused by the lack of technical understanding of the people responsible for eliciting requirements can be mitigated through supportive means. One such way is asking probe questions to identify architecturally significant requirements from software requirement specifications [6].

The twin peaks approach where requirements and architecture are developed in parallel [48], could be used to address many of the issues described above. While the twin peaks model alone does not eliminate them, it can provide a basis that act as a catalyst for prevention process together with above mentioned activities. The frequent cycles that moves between requirements and architecture, can improve the communication as well as the clarity of requirements as it requires both requirements and architecture to be clarified frequently. It also helps minimizing information distortion as the distorted information can be corrected before it cause a long team impact. The twin peak approach allows software teams to react quickly, thus accommodate changes in user requirements more easily and address requirement prioritization issues caused by a dynamic business environment [29]. However, the application of twin peak model should be carefully planned with consultation of all the teams involved in the project as variations in design cycles can affect project scheduling and synchronization.

## 10 | THREATS TO VALIDITY

According to Yin [65] a construct and external validity as well as reliability are necessary conditions that have to be taken into account when conducting case studies. Internal validity has to be considered when conducting exploratory case studies. Yin [65] suggests using multiple sources of evidence, establishing the chain of events and having key informants to review case study report as tactics for ensuring construct validity in case studies. Internal validity can be addressed for example through considering a rival explanation and using logical models.

Throughout the case study, various steps were taken to mitigate the threats to construct validity. The interview guide reviews done by multiple researchers and representatives of the case company. In addition, a pilot interview was conducted to get feedback from an expert to further improve the interview guide. Two researchers were participated in the interviews to make sure the questions and answers are interpreted correctly. At the beginning of each interview, key terms related to the study were defined and discussed to ensure a common vocabulary among researchers and interviewees. At the end of the each interview two researchers had a brief discussion to clarify any unclear situations. Maintaining chain of evidence is critical to the validity of the study. Hence the data analysis tool NVivo was used to record and analyze data throughout the study. The data as well as the thematic coding used to analyses data has been reviewed by multiple reviewers during the study.

The results of the data analysis were presented in the case company in a workshop with the participation of the researchers and the interviewees. The interviewees were given the opportunity give their feedback about results and point out if there were any inconsistencies. The possible ways to mitigate the issues and their practical implications were also discussed during the workshop. Later, the case study report was delivered to the case company representatives and the interviewees. All the recipients of the case study report were also requested to provide their feedback about the report.

Threats to internal validity must be taken into account when studying causal relationships. This study aimed to explore the challenges to software architecture posed by requirements volatility. Since software development is affected by several other factors, too, there is no clear causal relation between requirements volatility and software architecture challenges. The examples of these factors are technological changes and company strategies. However, this study addressed requirements volatility only.

External validity relates to the generalizability of results. Traditionally, it has been suggested that the generalizability of results from a single case study is rather poor [52]. However, the results of case studies may be extended to other cases that have common characteristics [52]. Seddon and Scheepers [53], suggest that generalisation of results can be done based on a single case study as long as 1) a sample is carefully analyses, 2) relevant factors, which are true in a sample can be argued being true in larger similar context and 3) researchers seeking to generalize results discuss their findings in relation to prior studies [53]. Even though that provides the opportunity to Considering the context of the study it is expected that similar finding can be drawn when the following characteristics are present: a) globally distributed software development teams, b) a company operating in a dynamic market, c) a large company structured as autonomous units and d) serving a diverse customer base. Threats to external validity were taken into account by collecting data that can be used to characterize the subjects and case context. Examples of these data are experience of the interviewees in their field and in the company, team sizes, organizational structures and roles and the responsibilities related to them.

The main limitation of this study regarding generalisation is that all the data were collected from a single case company. So there can be many context dependent factors that influence the software development of the case company as well as the outcome of the cases study. Since the case company consists of three different business lines and based on multiple development centers, cross analysis of data between the different business lines and development centers could have given opportunity to rule out some of the context dependent factors. However, since there are many common company wide policies, practices and cultural elements, such analyze does not help to generalise the findings in the same way as replicating the study in a different case company. Considering that replicating this study in another case company is the best possible way to improve the claim for generalisability. That was carefully considered during the reporting to make sure that there are sufficient information related to the context as well as the case study process to carry out a successful replication.

## 11 | LESSONS LEARNED

The case study process provide the researchers with several insights regarding conducting a case studies in large scale software development organization. Even though the management of the case company was convinced that the outcome of the study would be useful to improve the overall software development process in the case company, many of the interviewees were not convinced that the study would bring any benefit to them personally. So they were reluctant to spent time providing information. Despite the assurance from the researchers that the information collected will be anonymous, many of the interviewees were concerned about the privacy and specially about providing any negative information regarding the current process. It should be also noted that non disclosure agreements (NDAs) prevented many of them providing information to some of the questions and real examples. Overall, the researchers managed to improve the efficiency of data collection as the study progress and collect good quality data to analyze.

The semi structured interview based data collection requires interactive discussion. However, some of the interviewees provided short answers to the open ended questions and the researchers had to use different approaches to get the necessary information. The differences between the terminologies used by the researchers and the interviewees created communication challenges during the study. In some cases, lack of common understanding of a given term among the interviewees themselves, made it even more difficult to communicate. For example, since some of the interviewees didn't understand either 'non-functional requirements' or 'quality attributes', it should be elaborated further with an example before continuing the discussion. The researchers had to be cautious not to introduce bias or lead the interviewee to a certain direction during these clarifications.

## 12 | CONCLUSIONS AND FUTURE RESEARCH

This industrial case study was conducted to explore the challenges that requirements volatility poses to software architecture design. Fifteen software experts involved in software architecture design in various business units of a case company, which provides software products and services in a global market, were interviewed using a semi-structured interview as a guide.

This study revealed the visible factors of requirements volatility in each software team, as the well as their implications on software architecture. Furthermore, each teams' readiness to handle the implications of software volatility was also evaluated. Finally, the means to mitigate the identified implications and eliminate the possible factors were discussed. Poor communication, distorted information and external dependencies are the primary factors that imped software teams from choosing the optimal course of action to design software architecture, hence leading to architecture related issues such as inadequate documentation, inability to trace design rationale and increased complexity of the architecture.

As software engineering researchers are increasingly interested in the "twin peaks" of the software development: requirements and architecture design. This study provided empirical evidence about the relationship between them and how the process changes in one can affect another. The ultimate goal was to understand the complexity of the development environment and issues software teams face daily and thus propose feasible solutions for industry. This case study provides an example for practitioners how research may help to expose challenges, their reasons and impacts in the company. Software teams may consult the results of the case study to identify similarities and differences in their practices. This in turn helps to find improvement directions.

Requirement volatility often considered as an adversary. However, as highlighted in the agile software development principles [9], the changing nature of requirements can be utilized to improve the software products. While there is increased research interest on building architectures that can withstand on change [20, 1], we see a gap in understanding positive effect of requirements volatility on software architecture, especially in the industrial context. So we see empirical study on positive implications of requirements volatility on software architecture is good opportunity for future research.

20 | DASANAYAKE ET AL.[19] Ferreira S, Shunk D, Collofello J, Mackulak G, Dueck A. Reducing the risk of requirements volatility: findings from an empirical survey. Journal of Software Maintenance and Evolution: Research and Practice 2010;23(5):375–393.

[20] Ford N, Parsons R, Kua P. Building Evolutionary Architectures: Support Constant Change. Edition s, editor, O'Reilly Media; 2017.

[21] Fowler M, Is Design Dead?; 2004. http://martinfowler.com/articles/designDead.html.

[22] Fowler M, Highsmith J. The agile manifesto. Software Development 2001;9(8):28–35.

[23] Gil N, Beckman S. Design reuse and buffers in high-tech infrastructure development: A stakeholder perspective. IEEE Transactions on Engineering Management 2007;54(3):484–497.

[24] Greer D, Ruhe G. Software release planning: An evolutionary and iterative approach. Information and Software Technology 2004;46(4):243–253.

[25] Groher I, Weinreich R. A Study on Architectural Decision-Making in Context. In: 12th Working IEEE/IFIP Conference on Software Architecture (WICSA 2015); 2015. p. 11–20.

[26] Gubrium JF, Holstein JA. Handbook of Interview Research: Context and Method. Sage Publications; 2001.

[27] Hickey AM, Davis AM. Requirements elicitation and elicitation technique selection: Model for two knowledge-intensive software development processes. In: Proceedings of the 36th Annual Hawaii International Conference on System Sciences, HICSS 2003; 2003. .

[28] Hintersteiner J. Addressing changing customer needs by adapting design requirements. In: Proceedings of 1st International Conference on Axiomatic Design (ICAD 2000); 2000. p. 290–299.

[29] Huo M, Verner J, Zhu L, Babar Ma. Software quality and agile methods. Computer Software and Applications Conference, 2004 COMPSAC 2004 Proceedings of the 28th Annual International 2004;p. 520–525 vol.1.

[30] ISO/IEC/IEEE, Systems and software engineering — Architecture description; 2011.

[31] Jansen A, Bosch J. Software Architecture as a Set of Architectural Design Decisions. In: 5th Working IEEE/IFIP Conference on Software Architecture (WICSA'05); 2005. p. 109–120.

[32] Javed T, Maqsood Me, Durrani QS. A study to investigate the impact of requirements instability on software defects. ACM SIGSOFT Software Engineering Notes 2004;29(3):1.

[33] Jiang Tm, Coyner M. Software Process Disturbances. Computer Software and Applications Conference, 2000 COMPSAC 2000 The 24th Annual International 2000;p. 167–168.

[34] Klinger T, Tarr P, Wagstrom P, Williams C. An enterprise perspective on technical debt. In: Proceeding of the 2nd working on Managing technical debt (MTD '11); 2011. p. 35.

[35] Korkala M, Abrahamsson P, Kyllönen P. A Case Study on the Impact of Customer Communication on Defects in Agile Software Development. In: AGILE 2006 Conference (AGILE 2006); 2006. p. 76–88.

[36] Kotonya G, Sommerville I. Requirements Engineering : Processes and Techniques. Wiley; 1998.

[37] Kruchten P, Nord RL, Ozkaya I. Technical debt: From metaphor to theory and practice. IEEE Software 2012;29(6):18–21.

[38] Li Z, Liang P, Avgeriou P. Architectural Debt Management in Value-Oriented Architecting. In: Economics-Driven Software Architecture; 2014.p. 183–204.

[39] Loucopoulos P, Karakostas V. System Requirements Engineering. McGraw-Hill; 1995.

[40] Malaiya YK, Denton J. Requirements volatility and defect density. In: Proceedings 10th International Symposium on Software Reliability Engineering (Cat. No.PR00443); 1999. p. 285–294.

[41] Martini A, Bosch J, Chaudron M. Architecture technical debt: Understanding causes and a qualitative model. In: Proceedings - 40th Euromicro Conference Series on Software Engineering and Advanced Applications, SEAA 2014; 2014. p. 85–92.

[42] McGee S, Greer D. A Software Requirements Change Source Taxonomy. In: Fourth International Conference on Software Engineering Advances, 2009. ICSEA '09 Porto, Portugal: IEEE; 2009. .

[43] Miller JA, Ferrari R, Madhavji NH. An exploratory study of architectural effects on requirements decisions. In: Journal of Systems and Software, vol. 83; 2010. p. 2441–2455.

[44] Miyachi C. Agile software architecture. SEN - ACM SIGSOFT Software Engineering Notes 2011;36(2):1.